\documentclass[%
reprint,
amsmath,amssymb,
aps,
showkeys
]{revtex4-2}

\usepackage{graphicx}
\usepackage{dcolumn}
\usepackage{bm}
\usepackage{hyperref}
\hypersetup{colorlinks,allcolors=blue}

\usepackage{float}
\usepackage[braket, qm]{qcircuit}
\usepackage{graphicx}
\usepackage{multirow}
\usepackage{mathrsfs}
\usepackage{algorithm}
\usepackage{algpseudocode}

\begin{document}

\preprint{APS/123-QED}

\title{Full Quantum Process Tomography of a Universal Entangling Gate on an IBM's Quantum Computer}

\author{Muhammad AbuGhanem{$^{1,2,\star}$}}

\address{$^{1}$ Faculty of Science, Ain Shams University, Cairo, 11566, Egypt}
\address{$^{2}$ Zewail City of Science, Technology and Innovation, Giza, 12678, Egypt}

\email{gaa1nem@gmail.com}

\date{\today}

\begin{abstract}

Characterizing quantum dynamics is a cornerstone pursuit across quantum physics, quantum information science, and quantum computation. The precision of quantum gates in manipulating input basis states and their intricate superpositions is paramount. 
In this study, we conduct a thorough analysis of the SQSCZ gate, a universal two-qubit entangling gate, using real quantum hardware. This gate is a fusion of the square root of SWAP ($\sqrt{SWAP}$) and the square root of controlled-Z ($\sqrt{CZ}$) gates, serves as a foundational element for constructing universal gates, including the controlled-NOT gate.
we begin by explaining the theory behind quantum process tomography (QPT), exploring the \textit{Choi-Jamiolkowski} isomorphism or the Choi matrix representation of the quantum process, along with a QPT algorithm utilizing Choi representation.
Subsequently, we provide detailed insights into the experimental realization of the SQSCZ gate using a transmon-based superconducting qubit quantum computer. To comprehensively assess the gate's performance on a noisy intermediate-scale quantum (NISQ) computer, we conduct QPT experiments across diverse environments, employing both IBM Quantum's simulators and IBM Quantum's real quantum computer. 
Leveraging the Choi matrix in our QPT experiments allows for a comprehensive characterization of our quantum operations. 
Our analysis unveils commendable fidelities and noise properties of the SQSCZ gate, with process fidelities reaching $97.27098\%$ and $88.99383\%$, respectively. These findings hold promising implications for advancing both theoretical understanding and practical applications in the realm of quantum computation.

\end{abstract}


\maketitle

\section{Introduction}
\label{S:1}

Quantum information science has sparked a revolution, birthing a plethora of cutting-edge technologies and applications poised to redefine the future~\citep{NISQ23,QPT-hig1}. Among these groundbreaking domains, quantum cryptography~\citep{QPT-hig2}, quantum computation~\citep{Nilson}, and quantum sensing~\citep{QPT-hig4} shine as beacons of innovation, promising to usher in the next technological era~\citep{NISQ23,QPT-hig2,Nilson,QPT-hig4,art19,jiuzhang,Borealis5,Jiuzhang2.0,Zuchongzi}.

However, the realization of complex quantum machines and networks hinges upon a thorough understanding of their constituent components. Quantum process tomography (QPT) emerges as a powerful tool in this pursuit, enabling the meticulous reconstruction of component actions within quantum systems~\citep{chuang97,qpt}.

QPT's versatility is showcased in its application across diverse quantum physical systems. From the realm of photonic qubits~\citep{brien} and liquid-state NMR~\citep{QPT-hig7,QPT-hig8} to the intricacies of atoms in optical lattices~\citep{QPT-hig10} and trapped ions~\citep{QPT-hig11}, QPT has proven indispensable~\citep{QPT-hig7,QPT-hig8,brien,QPT-hig10,QPT-hig11}. 
Even in cutting-edge domains like continuous-variable quantum states~\citep{QPT-hig13}, solid-state qubits~\citep{QPT-hig12}, semiconductor quantum dot qubits~\citep{QPT-hig14}, and nonlinear optical systems~\citep{QPT-hig15}, with various practical implementations on NISQ computers~\citep{Entangling,Googles,Sycamore}. QPT continues to push the boundaries of characterization~\citep{QPT-hig12,QPT-hig13,QPT-hig14,QPT-hig15}.

Yet, the quest for comprehensive understanding extends beyond individual components to encompass entire quantum communication channels. Within the realms of quantum key distribution (QKD) and quantum communications~\citep{QCinternet3,IntegPhot6,IntegPhot7}, the characterization of quantum channels and components takes on paramount importance~\citep{QPT-hig16,QPT-hig17}. 
In this endeavor, quantum channels reveal themselves as conduits of potential both for seamless communication and clandestine interception. Categorized into optical fiber, line-of-sight free-space, and ground-to-satellite links, these channels serve as lifelines in the quantum realm~\citep{QPT-hig18,QPT-hig19,QPT-hig20}.

\medskip

Quantum process characterization, such as that of communication channels, serves as a crucial component in the establishment of quantum information systems. In quantum key distribution protocols, the level of noise present in the channel directly impacts the rate at which confidential bits are exchanged between authorized parties~\citep{QPT100-23,QPT100-24}. Specifically, tomographic protocols enable comprehensive reconstruction, thereby facilitating the thorough characterization of the channel~\citep{QPT100-25,QPT100-26,QPT-100}. 

\medskip

QPT constitutes a vital procedure aimed at validating quantum gates and identifying shortcomings in configurations and gate designs. 
The primary objective of QPT is to extract a comprehensive description of the dynamical map (quantum process) based on a series of well-designed experiments. 
In this paper, we perform QPT experiments utilizing the Choi matrix representation of quantum processes~\citep{Choi,Jamiołkowski}, in order to provide a complete characterization of the SQSCZ gate's performance on NISQ computers, the QPT analysis conducted across diverse environments, employing both IBM Quantum's simulators (\textit{qasm\_simulator}) and a real quantum computer (\textit{ibm\_perth})~\citep{IBM}.

\medskip

The subsequent sections of this paper are organized as follows: 
Section~\ref{sec:QPT} provides a succinct outline of the principles underlying quantum process tomography (QPT), emphasizing the \textit{Choi-Jamiolkowski} isomorphism or the Choi matrix representation of quantum processes, along with a QPT algorithm employing Choi representation. Following this theoretical foundation, Section~\ref{sec:SQSCZ} presents a concise overview of the two-qubit entangling gate, SQSCZ. In Section~\ref{sec:SQSCZ_IBM}, we delve into the experimental implementation of the SQSCZ gate utilizing a transmon-based superconducting qubit quantum computer. Building upon this experimental setup, Section~\ref{sec:QPT_EXP} encompasses our efforts to comprehensively assess the gate's performance on a NISQ computer through QPT experiments conducted across diverse environments, employing both IBM Quantum's simulators and real quantum computers. These QPT experiments utilized the Choi matrix to achieve a comprehensive characterization of our quantum operations. In Section~\ref{sec:Analysis}, we undertake tests and analyses to scrutinize gate fidelities and noise properties of the SQSCZ gate, paving the way for our ensuing discussion. Finally, we draw our conclusions in Section~\ref{SEC:Concl}.

\section{Quantum Process Tomography}\label{sec:QPT}

\subsection{Theory of QPT}

QPT aims to extract a complete description of a dynamical map ${\cal E}$ based on a series of well-designed experiments~\citep{qpt}. Standard QPT involves initializing system qubits in a suitable density matrix basis, subjecting these basis states to a specific process, and subsequently characterizing the resulting quantum states using a quantum state tomography algorithm~\citep{Sycamore}. % 
In this process, a density matrix ${\cal P}_{\text{in}}$ undergoes a linear transformation into another density matrix ${\cal P}_{\text{out}}$ by traversing through a general quantum operation ${\cal E}$, such that ${\cal P}_{\text{out}} = {\cal E}({\cal P}_{\text{in}})$. Here, ${\cal P}_{\text{in}}$ and ${\cal P}_{\text{out}}$ represent the input and output density matrices, respectively. The transformation 
%${\cal P}_{\text{in}} \xrightarrow{{\cal E}} {\cal P}_{\text{out}}$ 
${\cal E} : {\cal P}_{\text{in}} \longrightarrow  {\cal P}_{\text{out}}$
is fully characterized by a positive Hermitian matrix $\chi$, such that:

\begin{equation}
    {\cal P}_{\text{out}} \equiv {\cal E}({\cal P}_{\text{in}}) = \sum_{m,n} \chi_{mn} \widehat{W}_m {\cal P}_{\text{in}} {\widehat{W}}^{\dag}_{n},
\end{equation}

Here, ${\cal E}$ represents a completely positive quantum dynamical map, which illustrates the system's behavior on any d-dimensional input state ${\cal P}_{\text{in}}$. This map is commonly known as a quantum process.
The trace-preserving positive Hermitian matrix $\chi_{mn}$, with dimensions $d^2 \times d^2$, known as the ``process matrix," 
fully and distinctly characterizes the operation of the process ${\cal E}$.  
In this context, the operators $\widehat{W}_m$ constitute a complete basis, usually represented by the Pauli operators or, in higher dimensions, the Gell-Mann operators. Also, the dimensions of the $\widehat{W}_m$ matrices match those of density operators. It's noteworthy that the dimensions of the $\widehat{W}_m$ matrices match those of density operators. For instance, in scenarios involving $K$ qubits, where $d=2^K$, there exist $4^K$ operators $\widehat{W}_n$, resulting in a process matrix dimension of $4^K \times 4^K$.

\medskip

Various mathematical representations exist for a completely positive (${\cal E} \otimes \mathbb{I} \ge zero$) and trace-preserving ($\rm{tr} ({\cal E}({\cal P}))= \rm{tr}({\cal P})$) (CPTP) map~\citep{Nilson,QPT-47}, including Choi representation~\citep{Choi,Jamiołkowski}, the Kraus representation~\citep{kraus83},  and Pauli-Transfer-Matrix (PTM) representation~\citep{PTM1}. For a summary of these representations, please refer to reference~\citep{QPT-47}. 
A standalone version of QPT, which does not necessitate quantum state tomography (QST), is outlined in~\citep{QPTGambetta}. The process of QPT for a high-dimensional quantum communication channel is elucidated in~\citep{QPT-100}. 
The methodology of QPT employing completely positive and trace-preserving projection is examined in~\citep{QPT-101}. QPT employing unsupervised learning and tensor networks is demonstrated in~\citep{QPT-102}. 
For a comprehensive understanding of QPT theory, detailed descriptions are available in \citep{kraus83,chuang97,qpt,chell03,brien}. 
In our QPT experiments, we employed the \textit{Choi-Jamiolkowski} isomorphism~\citep{Choi,Jamiołkowski}, also known as the Choi matrix, to achieve a comprehensive characterization of our quantum operations. This approach %, as described by Choi and Jamiołkowski, 
allowed us to effectively represent and analyze the behavior of our quantum processes.

\subsection{The Choi representation}

An alternative and convenient conceptualization of procedures is presented through the \textit{Choi-Jamiolokowski} isomorphism~\citep{Choi,Jamiołkowski}, asserting that any entirely positive mapping can be elucidated as an operator situated within a $d^2$-dimensional Hilbert space $ {\cal H}_\mathrm{in} \otimes {\cal H}_\mathrm{out}$. This operator, denoted as the Choi matrix 
${\cal C}_{\cal E}$, is constructed as the outcome of the channel's operation on one component of a maximally entangled state.

\medskip

Therefore, a quantum channel ${\cal E}$ can be fully characterized by a unique bipartite matrix known as the Choi matrix ${\cal C}_{\cal E}\in \mathscr{L}({\cal H}_\mathrm{in} \otimes {\cal H}_\mathrm{out} )$, which is derived from the \textit{Choi-Jamiolkowski} isomorphism~\citep{Choi,Jamiołkowski}. 
Now, if the set of kets $\left\{ \ket{0}, \ket{1}, \dots, \ket{j-2},\ket{j-1}\right\}$ forms an orthonormal set of basis of both the input ${\cal H}_\mathrm{in}$ and the output ${\cal H}_\mathrm{out}$ Hilbert spaces, then the Choi operator (matrix) ${\cal C}_{\cal E}$ is defined as follows:

\begin{equation}
    {\cal C}_{\cal E} = \sum_{k,m=0}^{j-1}\ket{k}\bra{m}\otimes {\cal E} \left(\ket{k}\bra{m}\right).
\end{equation}

\noindent
Using this Choi matrix representation, the evolution of a given quantum state ${\cal P}_{\text{in}}$ can be described by:

\begin{equation}
    {\cal E}({\cal P}_{\text{in}}) = \textrm{Tr}^\pounds_{{\cal H}_{in}} \left[ ({\cal P}_{\text{in}}^T \otimes \mathbb{I}) \, {\cal C}_{\cal E} \right],
\end{equation}

\noindent
here, the symbol $\textrm{Tr}^\pounds$ represents the partial trace over the input state’s
Hilbert space, %such that $ \textrm{Tr}^\pounds_{{\cal H}_{in}} [K\otimes M] = \textrm{Tr}[K]\, M$, 
with $\mathbb{I}$ defined as the (d-dimensional) identity operator, and $T$ signifies the matrix transposition.

\medskip

When preparing the quantum system in an initial state ${\cal P}_k$ and subsequently performing a measurement, the probability of obtaining the measurement outcome ($km$) for a projector $\Xi$  can be written as

\begin{equation}\label{pp}
    p_{km} = \textrm{Tr} \left[ ({\cal P}_k^T \otimes \Xi_m) \, {\cal C}_{\cal E} \right]
\end{equation}

The expression provided in Eq. (\ref{pp}) reveals that measuring the outcome of the quantum channel ${\cal E}$ acting on an initial state ${\cal P}_k$ using a projector $\Xi$ is similar to measuring the Choi matrix (${\cal C}_{\cal E}$) as a quantum state within the space $\mathscr{L}({\cal H}_\mathrm{in} \otimes {\cal H}_\mathrm{out} )$ with a measurement operator $({\cal P}^T \otimes \Xi)$.

\medskip

In contrast to the Kraus representation~\citep{kraus83}, the Choi matrix associated with a quantum channel possesses uniqueness. Consequently, QPT essentially entails adjusting the matrix elements of ${\cal C}_{\cal E}$ to align with the data. The data set comprises a distinct collection of prepared input states introduced to the channel and a corresponding set of measurements on the resulting output states. Specifically, a collection of input states and measurements attains the status of being  informationally-complete if the inputs and the measurement operators collectively span the entire input and output Hilbert spaces of the quantum channel, respectively.

\subsection{QPT algorithm with the Choi representation}

The conventional methodology for conducting QPT entails parameterizing the Choi matrix, 
typically represented by a $4^K \times 4^K$ matrix, and deducing its constituents through the resolution of the maximum likelihood estimation problem~\citep{QPT-22}. This approach encounters two primary constraints. Firstly, it necessitates the parameterization of the entire Choi matrix, a task that escalates exponentially in complexity with the increase in the number of qubits. Secondly, to attain a high-fidelity alignment, the exhaustive informationally-complete set (ICS) of input states and measurements is indispensable, a requirement that also grows exponentially with $K$. Consequently, the practice of full QPT has remained confined to exceedingly diminutive system dimensions. 
To date, owing to these constraints, QPT has been applied experimentally to a maximum of 3 qubits
~\citep{brien,QPT-hig11,QPT-MaxQ25,QPT-MaxQ26,QPT-MaxQ27,QPT-MaxQ28,QPT-MaxQ29,QPT-MaxQ30,Googles,Sycamore,Entangling,Entangling0}.

To conduct quantum process tomography (QPT) experiments using the Choi matrix, the following steps are followed:

\begin{itemize}
    \item Step 1: Prepare a set of initial states, $\{ {\cal P}_1, {\cal P}_2, \dots, {\cal P}_{s}\}$.
    \item Step 2: Construct a set of projectors $\{P_1, P_2,\dots,P_t\}$ such that, for the state space $\mathscr{L}({\cal H}_\mathrm{in} \otimes {\cal H}_\mathrm{out} )$, the set of all projectors $\Xi_{i,j}=({\cal P}^T_i \otimes P_j)$ forms a tomographically complete set.
    \item Step 3: Initialize the system to ${\cal P}_i$. Utilize a quantum computer or a qasm simulator to apply the channel ${\cal E}$ and measure with $P_j$.
    \item Step 4: Acquire measurements of ${\cal C}_{\cal E}$ with $\Xi_{ij}$.
    \item Step 5: Pass the experimental results and the specification of $\Xi$ through a QST algorithm.
    \item Step 6: Utilize the measurement outcomes to construct the Choi matrix ${\cal C}_{\cal E}$.
\end{itemize}

\section{An Overview of the SQSCZ Gate}\label{sec:SQSCZ}

\begin{figure*}
    \centering
    \includegraphics[width=0.95\textwidth]{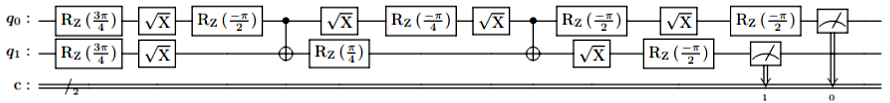}
\caption{A quantum circuit representation illustrating the de-construction of the SQSCZ gate into individual single-qubit operations and CNOT operations. The deconstruction of the SQSCZ gate involves a combination of single-qubit operations, namely rotation gates denoted as $\mathrm{R_Z}(.)$, and the square root of the NOT gate ($\mathrm{\sqrt{X}}$), along with the utilization of only two controlled-NOT (CNOT) gates.}
    \label{fig:SQSCZ_DECOM}
\end{figure*}

\begin{figure*}
    \centering
    \includegraphics[width=0.95\textwidth]{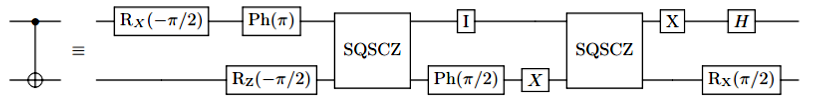}
\caption{A quantum circuit that serves as an alternative to the controlled-NOT (CNOT) gate is devised, leveraging the SQSCZ gate. This CNOT gate configuration entails the integration of solely two SQSCZ gates, supplemented by a sequence of single-qubit gates. These single-qubit operations encompass the $X$ gate, rotation gates ($\mathrm{R}_X, \mathrm{R}_Z$), and phase gates ($\mathrm{Ph}$).} 
 \label{fig:cnot_CIRC}
\end{figure*}

\begin{figure}
    \centering
    \includegraphics[width=0.49\textwidth]{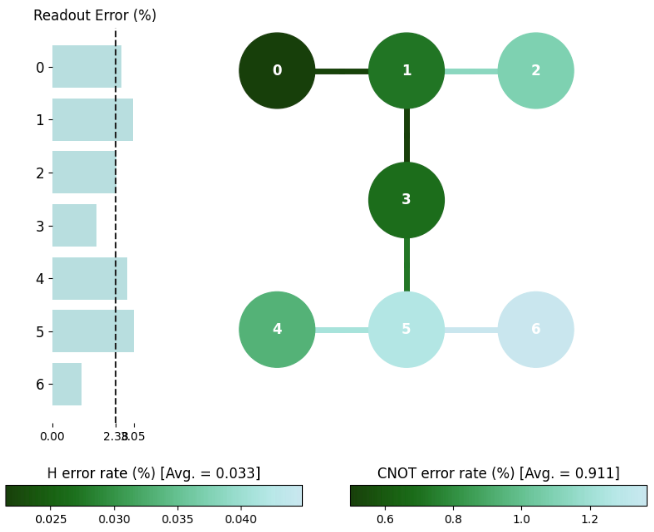}
    \caption{
The error map of the IBM Quantum machine "\textit{ibm\_perth}" and its corresponding layout. This processor, identified as the Falcon r5.11H version 1.2.8, comprises seven superconducting transmon qubits, boasting a quantum volume (QV) of 32. The basis gates available on this quantum computer include CNOT, ID, RZ, SX, and X gates.
Specific error metrics of this quantum computer, revealing a median CNOT error of $8.690 \times 10^{-3}$, a median SX error of $2.860 \times 10^{-4}$, and a median readout error of $2.930 \times 10^{-2}$. Additionally, the median values for the coherence times T1 and T2 were found to be $168.65$ microseconds ($\mu$s) and $132.51$ microseconds ($\mu$s), respectively.
Furthermore, the processor is capable of executing up to 100 quantum circuits, with a maximum of $20,000$ shots per circuit.}
    \label{fig:error_rate}
\end{figure}

\begin{figure}
    \centering
    \includegraphics[width=0.5\textwidth]{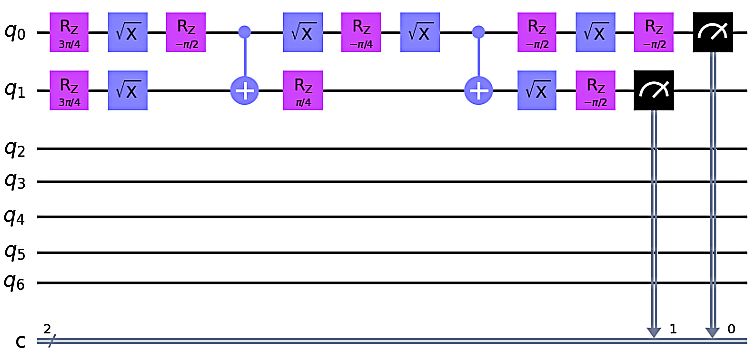}
    \caption{
An analogous quantum circuit representing the SQSCZ two-qubit entangling gate is presented. The gate is deconstructed utilizing single-qubit gates, employing merely two Controlled-NOT (CNOT) operations. The execution of this quantum circuit is repated for 7,168 shots on the seven-qubit IBM quantum computer, named \textit{ibm\_perth}.}
    \label{fig:sqscz0}
\end{figure}

\begin{figure}
    \centering
    \includegraphics[width=0.5\textwidth]{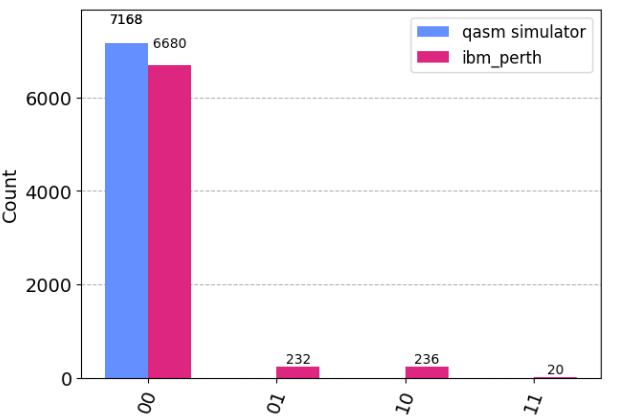}
    \caption{The measurement outcomes stemming from the practical implementation of the two-qubit SQSCZ gate were meticulously analyzed. These results were derived following the execution of the experiment with $7,168$ shots, both on a qasm simulator and a real quantum computer, specifically the IBM Quantum device known as "\textit{ibm\_perth}".  The presence of small probabilities in the outcomes can be attributed to noise inherent in quantum computation. The provided data represents the frequencies obtained for each of the two-qubit basis states.}
    \label{fig:SQSCZIm}
\end{figure}

\begin{table*}
\caption{\label{table:perth}The general operational efficacy of the seven-qubit IBM's Quantum superconducting quantum computer, known as \textit{ibm\_perth}. Key parameters considered include 
coherence times $T1$ and $T2$,
qubit frequency ($\omega/2\pi$), 
anharmonicity ($\gamma/2\pi$), readout length ($\delta$), readout assignment error ($\xi$), as well as the probabilities of bit flip ($\zeta$ and $\eta$), alongside single-qubit Pauli X/(SX) error. These metrics were accessed on September 23, 2023.}
\centering 
\begin{ruledtabular}
\begin{tabular}{c c c c c c c c c c}
$qubit~[i]$ & $T1(\mu s)$ &  $T2(\mu s)$ & $  \omega/2\pi \textrm{(GHz)}$& $\gamma/2\pi \textrm{(GHz)}$&$  \xi(R)$&  $ \zeta (\%)$ &$ \eta (\%)$&  $ \delta$(ns)& Pauli X/(SX) error  \\
\hline \hline
%Qubit &T1 (us)	&T2 (us)	&Frequency (GHz)	&Anharmonicity (GHz)	&Readout assignment error 	&Prob meas0 prep1 	&Prob meas1 prep0 	&Readout length (ns)	&Pauli / $\sqrt{X}$ (sx) error \\
$qubit~[0]$	&112.21714928	&89.95798047	&5.15755952 	&-0.341524517	&0.0281	 &0.0292	&0.027	&721.7777778	&0.00100025 \\
$qubit~[1]$	&143.97956535	&54.15814934	&5.03354584 	&-0.344368702	&0.0304	 &0.032	    &0.0288	&721.7777778	&0.00028598 \\
$qubit~[2]$	&214.03073768	&95.99622158	&4.86265682 	&-0.347272472	&0.0338	 &0.0272	&0.0404	&721.7777778	&0.00026989 \\
$qubit~[3]$	&168.64915374	&227.9672909	&5.12510400 	&-0.340441899	&0.0161	 &0.0186	&0.0136	&721.7777778	&0.00024909 \\
$qubit~[4]$	&169.53935454	&132.5137580	&5.15920959 	&-0.333366537	&0.0293	 &0.0286	&0.03	&721.7777778	&0.00030961 \\
$qubit~[5]$	&148.76564849   &142.0891481	&4.97860983 	&-0.346022031	&0.03	 &0.0344	&0.0256	&721.7777778	&0.00034369 \\
$qubit~[6]$	&188.84148933	&231.3818681	&5.15663665 	&-0.340454390	&0.0114	 &0.0126	&0.0102	&721.7777778	&0.00026888 \\
\end{tabular}
\end{ruledtabular}
\end{table*}

\begin{table}[]
\begin{ruledtabular}
\caption{\label{table:cnot} The error rates of controlled-NOT gates between distinct qubits ($qubit~[i]$) while operating on the "$ibm\_perth$" quantum computer, as accessed on September 23, 2023.}
\centering
\begin{tabular}{c c c}
%\hline \hline
$qubit$[index] &The coupling map & Error rates  \\
\hline \hline  
$qubit~[0]$ & CX: $qubit~[0]-qubit~[1]$  &0.00514 \\
\hline
\multirow{3}{*}{$qubit~[1]$}&
 CX: $qubit~[1]-qubit~[2]$  &0.01136  \\
&CX: $qubit~[1]-qubit~[3]$  &0.00499  \\
&CX: $qubit~[1]-qubit~[0]$  &0.00514  \\
\hline
$qubit~[2]$&
CX: $qubit~[2]-qubit~[1]$  &0.01136  \\
\hline
\multirow{2}{*}{$qubit~[3]$}&
CX:  $qubit~[3]-qubit~[1]$  &0.00499  \\
&CX: $qubit~[3]-qubit~[5]$  &0.00723  \\
\hline
$qubit~[4]$&
CX: $qubit~[4]-qubit~[5]$  &0.01226  \\
\hline
\multirow{3}{*}{$qubit~[5]$}&
CX:  $qubit~[5]-qubit~[4]$  &0.01226  \\
&CX: $qubit~[5]-qubit~[3]$  &0.00723  \\
&CX: $qubit~[5]-qubit~[6]$  &0.01367  \\ 
\hline
$qubit~[6]$&
CX: $qubit~[6]-qubit~[5]$  &0.01367  \\
%\hline \hline
    \end{tabular}
    \label{tab:my_label}
\end{ruledtabular}
\end{table}

A paradigm shift toward quantum computing~\citep{DiVincenzo,Nilson,arch02,qc,BA52} promises substantial practical implications, offering transformative potential for computational methodologies, programming techniques, and complexity paradigms~\citep{NISQ23,art19,jiuzhang,Borealis5,Jiuzhang2.0,Zuchongzi}. However, the realization of a functional quantum computer critically hinges upon satisfying specific criteria, meticulously outlined by DiVincenzo~\citep{DiVincenzo}. Among these criteria, a pivotal challenge lies in the physical realization of quantum gates operating between qubits~\citep{BA52,br83,IBM22,m14,algorithms}.

\medskip

Universal quantum gates play an indispensable role in expediting efficient algorithm execution on quantum computers, providing both rapid and fault-tolerant manipulations. Serving as foundational components, these gates are necessary for encoding intricate algorithms and operations within the  large-scale quantum computing systems~\citep{nisqQC11}.

The pursuit of high-fidelity two-qubit gates poses a formidable challenge, particularly in experimental settings encompassing diverse quantum systems. In the quest for a universal quantum computerr~\citep{NISQ23,qc,Nilson}, a fundamental suite of quantum gates, including  two-qubit entangling gates~\citep{br83,yao19,Entangling,Entangling0}, is essential. The effective integration of these gates signifies a crucial milestone in the progression toward practical quantum computing systems~\citep{MC5}. 
In this regard, a scheme has been introduced~\citep{SQSCZ,AbuGhanemMSc} to realize a novel two-qubit entangling gate within Josephson junctions superconducting qubits in~\citep{JosephsonSC}. This gate, known as SQSCZ, transforms the basis states as follows:

\begin{equation}\label{s11}
\begin{aligned}
&\ket{0}_A\otimes\ket{0}_B\rightarrow \ket{0}_A\otimes\ket{0}_B, 
\\&
\ket{0}_A\otimes\ket{1}_B\rightarrow \frac{1}{2}(1+i)\ket{0}_A\otimes\ket{1}_B +\frac{1}{2}(1-i) \ket{1}_A\otimes\ket{0}_B, 
 \\&
\ket{1}_A\otimes\ket{0}_B\rightarrow \frac{1}{2}(1-i)\ket{0}_A\otimes\ket{1}_B +\frac{1}{2}(1+i) \ket{1}_A\otimes\ket{0}_B, %\textrm{and}
\\&
\ket{1}_A\otimes\ket{1}_B\rightarrow i\ket{1}_A\otimes\ket{1}_B 
\end{aligned}
\end{equation}

Here, the representation is in the 2-qubit basis $ \left\{\ket{0}_A\otimes\ket{0}_B,\, \ket{0}_A\otimes\ket{1}_B,\, \ket{1}_A\otimes\ket{0}_B,\, \ket{1}_A\otimes\ket{1}_B\right\}$, and the subscript $A$ signifies qubit A.
The matrix operator corresponding to SQSCZ reads,
\begin{equation}\label{s11}
\begin{aligned}
\hat{U}^{SQSCZ}
&=
\begin{pmatrix}
1& 0 & 0 & 0 \\
0& \frac{1}{2}(1+i) & \frac{1}{2} (1-i) & 0 \\
0& \frac{1}{2} (1-i) & \frac{1}{2} (1+i)& 0 \\
0& 0 & 0 &  i
\end{pmatrix}
\end{aligned}
\end{equation}

\medskip

The SQSCZ-gate is a fusion of the 
$\sqrt{SWAP}$-gate and the $\sqrt{CZ}$-gate, which can be presented in
the form $\hat{U}^{\sqrt{SWAP}}\quad \hat{U}^{\sqrt{CZ}}=
\hat{U}^{\sqrt{CZ}}\quad \hat{U}^{\sqrt{SWAP}}$, where
\begin{equation}\label{}
\hat{U}^{\sqrt{SWAP}}=
\begin{pmatrix}
1& 0     & 0     & 0 \\
0& \frac{1}{2}(1+i) & \frac{1}{2}(1-i) & 0 \\
0& \frac{1}{2}(1-i) & \frac{1}{2}(1+i) & 0 \\
0& 0     & 0     &  1
\end{pmatrix}
\end{equation} 
and 
\begin{equation}\label{}
\hat{U}^{\sqrt{CZ}}=
\begin{pmatrix}
1& 0 & 0 & 0 \\
0& 1 & 0 & 0 \\
0& 0 & 1 & 0 \\
0& 0 & 0 & i
\end{pmatrix}
\end{equation}

are the $\sqrt{SWAP}$ and $\sqrt{CZ}$ gates expressed within the identical basis as Eq. (\ref{s11}).

\medskip

A quantum circuit representing the SQSCZ gate is depicted in Figure~\ref{fig:SQSCZ_DECOM}. This schematic illustrates a circuit decomposition that offers an alternative formulation of the SQSCZ two-qubit gate in terms of CNOTs gates. This decomposition strategy employs a combination of single-qubit operations, specifically rotation gates denoted as $\mathrm{R_Z}$, along with the square root of NOT gate ($\mathrm{\sqrt{X}}$), facilitated by the inclusion of only two controlled-NOT (CNOT) gates.
Furthermore, the universal controlled-NOT gate (CNOT) can be synthesized from the SQSCZ gate and single-qubit rotations through various combinations. One such combination can be expressed as: 
$[R_X(-\pi/2)Ph(\pi)\otimes R_Z(-\pi/2)] \quad \hat{U}^{SQSCZ}\quad [I\otimes Ph(\pi/2)\,X] \quad\hat{U}^{SQSCZ}\quad [X H\otimes R_X(\pi/2)]$. 
Here, the rotations are conducted with angles specified in radians. A quantum circuit equivalent to the controlled-not (CNOT) gate, formulated in terms of the SQSCZ gate, is introduced in Figure~\ref{fig:cnot_CIRC}.

\section{The Practical Implementation of the SQSCZ Gate}\label{sec:SQSCZ_IBM}

In this study, for the experimental implementation of our two-qubit quantum gate, we utilized the IBM Quantum's quantum computer known as \textit{ibm\_perth}. At the core of this quantum computing machine lies a quantum processor of type Falcon r5.11H version 1.2.8, featuring seven superconducting transmon qubits arranged according to the configuration depicted in Figure~\ref{fig:error_rate}. The available basis gates on this quantum machine include CNOT, ID, RZ, SX, and X gates. Noteworthy error metrics for this quantum computer include a median CNOT error of $8.690 \times 10^{-3}$, a median SX error of $2.860 \times 10^{-4}$, and a median readout error of $2.930 \times 10^{-2}$. Additionally, median values for the coherence times T1 and T2 were determined to be $168.65$ microseconds ($\mu$s) and $132.51$ microseconds ($\mu$s), respectively. Furthermore, the processor boasts the capability to execute up to 100 quantum circuits, with a maximum of $20,000$ shots per circuit. 
The general system characteristics and qubit properties of the \textit{ibm\_perth} quantum computer are listed in Tabel~\ref{table:perth}, whereas, the error rates of controlled-NOT gates between distinct qubits is presented in Tabel~\ref{table:cnot}.

\medskip

The quantum circuit representing the SQSCZ gate is executed for $7,186$ shots on both the \textit{qasm\_simulator} and the \textit{ibm\_perth} quantum computer, see Figure~\ref{fig:sqscz0}. 
The \textit{qasm\_simulator} is designed to emulate quantum circuits under various conditions, including both pristine and noisy environments. It supports a range of simulation methodologies, each adaptable to meet the specific needs of circuits and noise models. While the \textit{qasm\_simulator} replicates an ideal quantum computing setting, it also provides the capability to mimic the characteristics of real-world noisy quantum computers. For more in-depth information on IBM Quantum's simulators, readers are referred to~\citep{IBM}.

\medskip

The results of the measurements indicate that: On the \textit{qasm\_simulator}, out of the $7,186$ shots, the classical register returns the state $\ket{00}$ for the entirety of the measurements, while yielding zero occurrences for the computational basis states $\ket{01}$, $\ket{10}$, and $\ket{11}$. Conversely, in the experimental implementation on the \textit{ibm\_perth} quantum computer, the measurement outcomes are as follows: $\ket{00}$ is obtained in $6,680$ instances, $\ket{01}$ in $232$ instances, $\ket{10}$ in $236$ instances, and $\ket{11}$ in $20$ instances. The measurement outcomes from the actual execution of the two-qubit gate on both the \textit{qasm\_simulator} and the quantum computer are illustrated in Figure~\ref{fig:SQSCZIm}. Qubit properties and system performance during this experimental implementations are presented in Tabel~\ref{table:SQSCZ}.

\section{The QPT Experiments}\label{sec:QPT_EXP}

%QPT MATRIX 
\begin{figure*}
    \centering
    \includegraphics[width=0.75\textwidth]{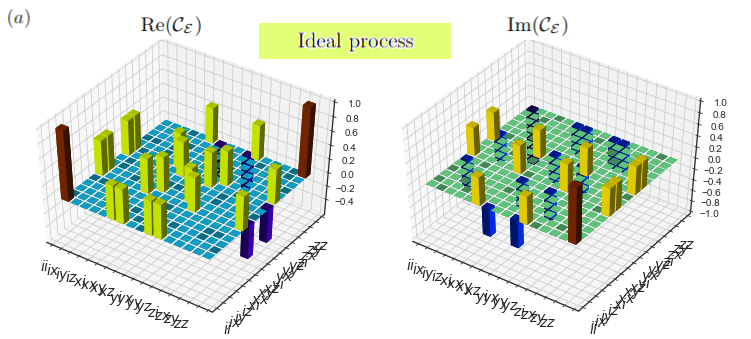}
    \includegraphics[width=0.75\textwidth]{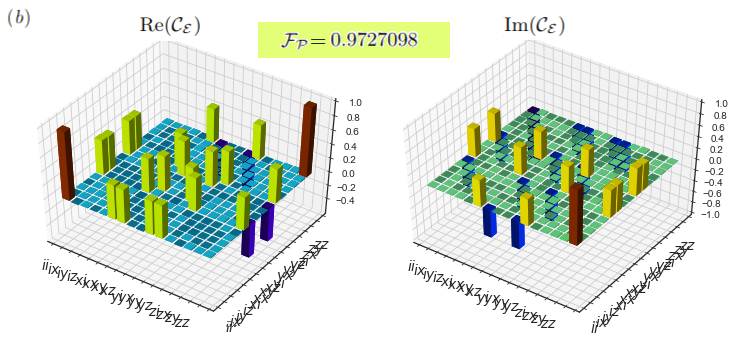}\\
    \includegraphics[width=0.75\textwidth]{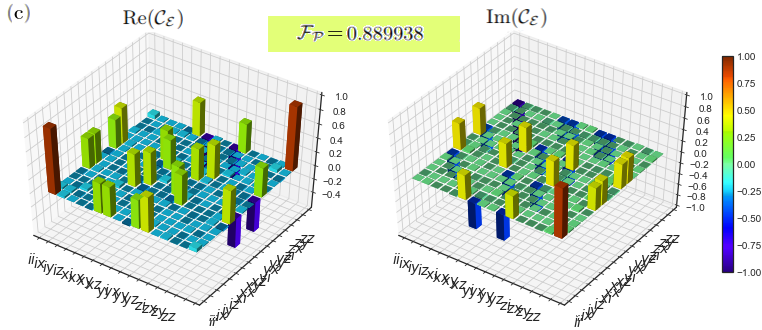}
    \caption{
    The real ($\mathrm{Re}({\cal C}_{\cal E})$) and imaginary ($\mathrm{Im}({\cal C}_{\cal E})$) components of the Choi matrices, reconstructed the QPT experiments for the two-qubit SQSCZ gate, are depicted alongside the final results for the measurement outcomes across all circuits.
(a) Represents the ideal Choi matrix (theoretical expectations).
(b) Illustrates the results of QPT experiments conducted on an IBM Quantum simulator (\textit{qasm\_simulator}).
(c) Depicts the outcomes of QPT experiments carried out on a real IBM quantum computer, \textit{ibm\_perth}. The QPT experiments were repeated 11,000 times on the IBM Quantum's \textit{qasm\_simulator} and for $4,000$ times in IBM's Quantum compter \textit{ibm\_perth}, yielding process fidelities ${\cal F_P}$ of $97.27098\%$ and $88.9938\%$, respectively. The total runtime on the \textit{ibm\_perth} machine amounted to 2 minutes and 40 seconds.}
    \label{fig:QPTMatrix}
\end{figure*}

% HINTON
\begin{figure*}
    \centering
    \includegraphics[width=0.3\textwidth]{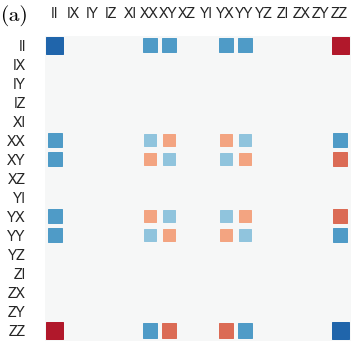}
    \includegraphics[width=0.3\textwidth]{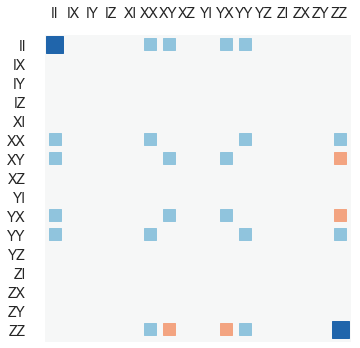}
    \includegraphics[width=0.35\textwidth]{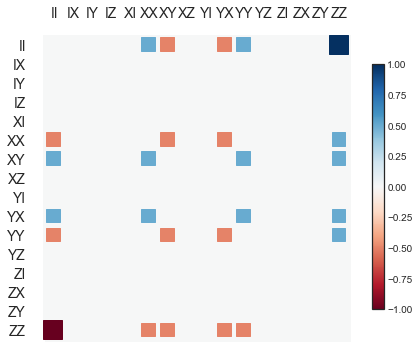}\\
    \vspace{0.5cm}
    \includegraphics[width=0.3\textwidth]{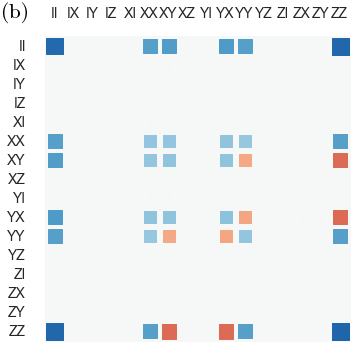}
    \includegraphics[width=0.3\textwidth]{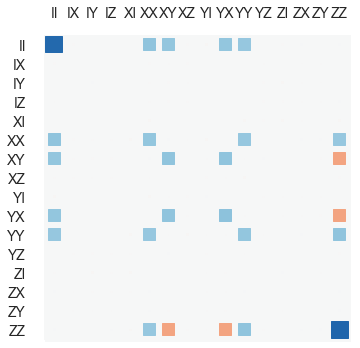}
    \includegraphics[width=0.35\textwidth]{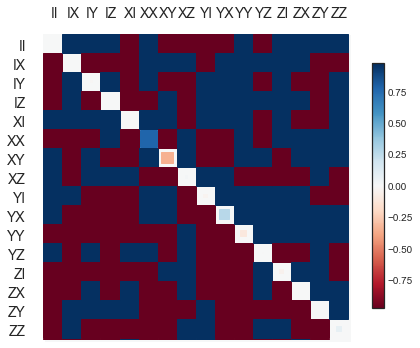}\\
    \vspace{0.5cm}
    \includegraphics[width=0.3\textwidth]{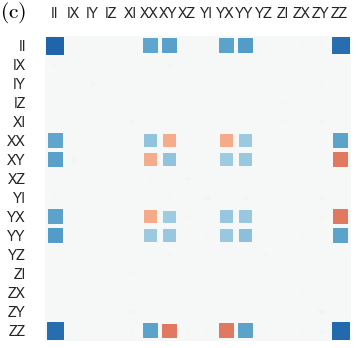}
    \includegraphics[width=0.3\textwidth]{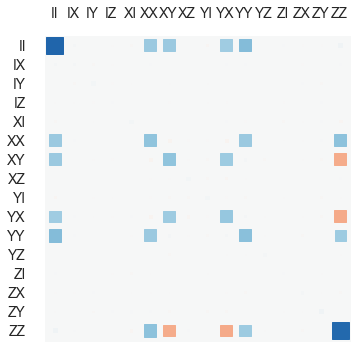}
    \includegraphics[width=0.35\textwidth]{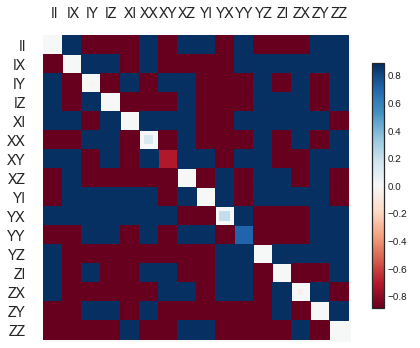}
    \caption{
Quantum process tomography (QPT) of the two-qubit SQSCZ gate is conducted on the sixteen two-qubit basis states in the Pauli basis: $II, IX, IY, IZ, XI, XX, XY, XZ, YI, YX, YY, YZ, ZI, ZX, ZY$, and $ZZ$, visualized as Hinton diagrams.
(a) Illustrates visualizations of the ideal process matrix of the two-qubit SQSCZ gate, devoid of effects arising from decoherence.
(b) Depicts the results of two-qubit process tomography for the SQSCZ gate executed on an IBM Quantum simulator (\textit{qasm\_simulator}) with 11,000 shots.
(c) Displays QPT experiments performed for 4,000 shots on a real IBM quantum computer, \textit{ibm\_perth}.
The left panel illustrates the combined parts, while the middle and right plots show the real and imaginary components, respectively.
We demonstrate process fidelities of $97.27098\%$ and $88.9938\%$ on the IBM Quantum simulator and the IBM quantum computer \textit{ibm\_perth}, respectively.
}
    \label{fig:QPTHinton}
\end{figure*}

% PARAMETERS
\begin{figure*}
    \centering
    \includegraphics[width=0.32\textwidth]{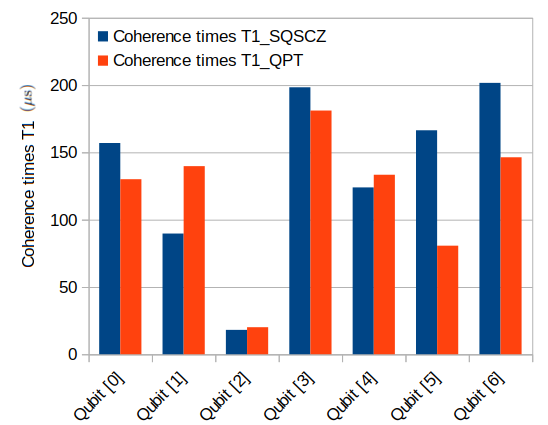}
    \includegraphics[width=0.32\textwidth]{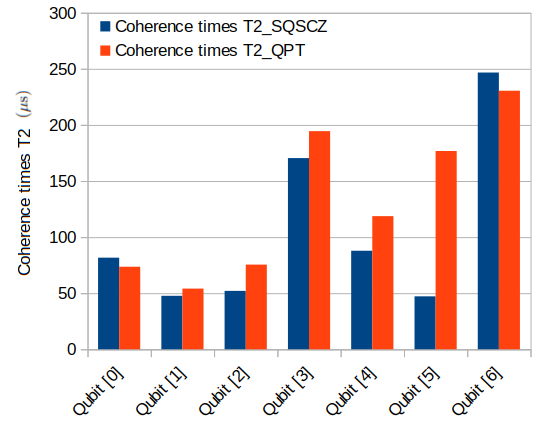}
    \includegraphics[width=0.32\textwidth]{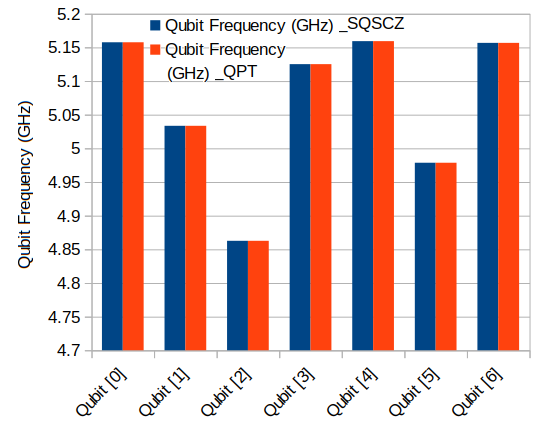}\\
    \includegraphics[width=0.32\textwidth]{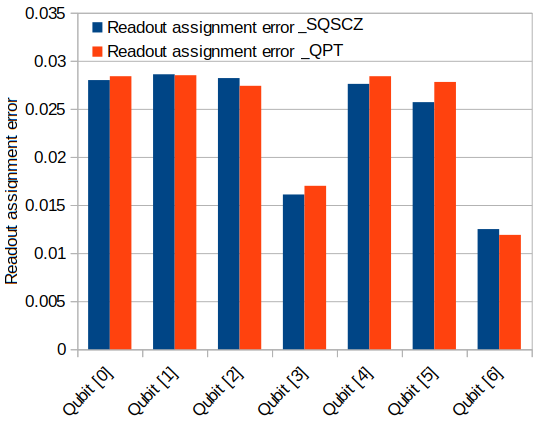}
    \includegraphics[width=0.32\textwidth]{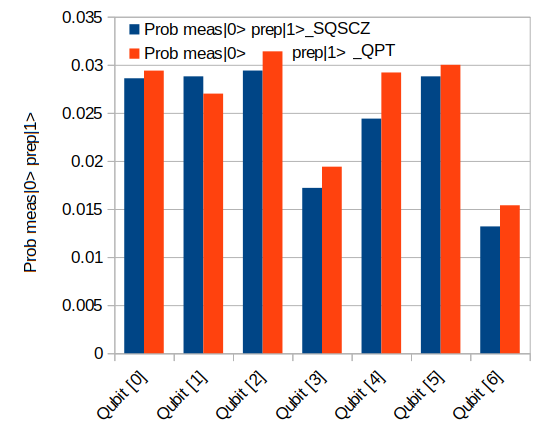}
    \includegraphics[width=0.32\textwidth]{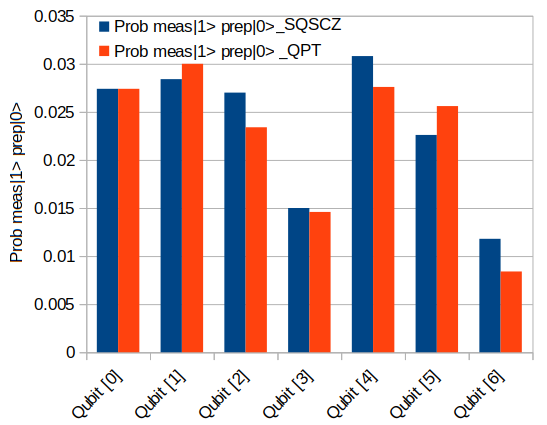}\\
    \includegraphics[width=0.32\textwidth]{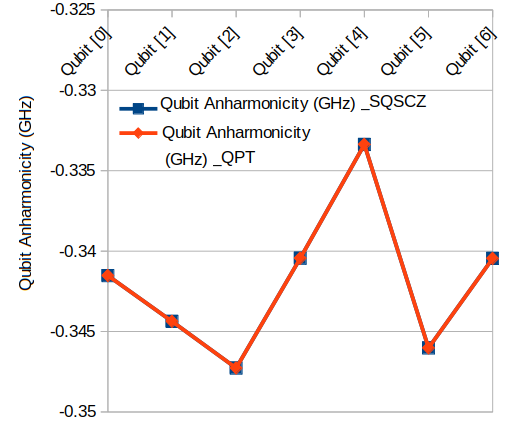}
    \includegraphics[width=0.32\textwidth]{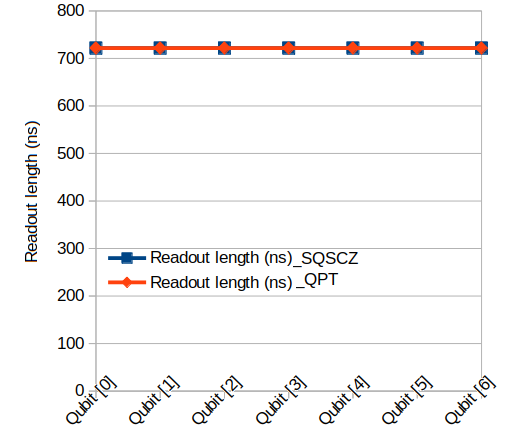}
    \includegraphics[width=0.32\textwidth]{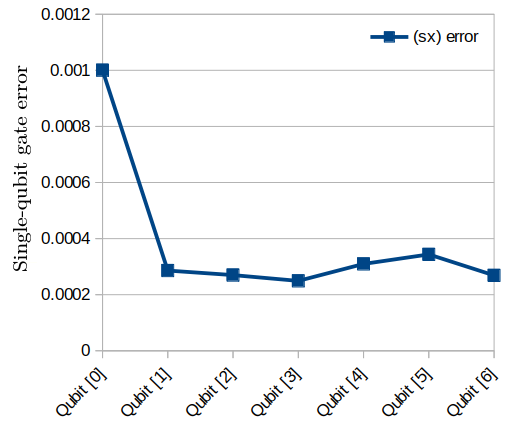}
    \caption{
Performance metrics, distribution of parameters, and readout error analysis for individual qubits on the IBM's quantum processor \textit{ibm\_perth} during the execution of the two-qubit SQSCZ quantum circuit and quantum process tomography experiments pertaining to the SQSCZ gate.
(a) T1 (microseconds), 
(b) T2 (microseconds), 
(c) Qubit Frequency (gigahertz), 
(e) Readout assignment error, 
(f) Probability of measuring $\ket{0}$ when preparing $\ket{1}$, and 
(g) probability of measuring $\ket{1}$ when preparing $\ket{10}$. 
(d) Qubit Anharmonicity (gigahertz), 
(h)and Readout length (nanoseconds).
 SX gate error.
}
    \label{fig:parameters}
\end{figure*}

In this section, to comprehensively characterize our quantum gate and identify potential flaws in configuration and gate design, we perform QPT experiments for the SQSCZ gate. These quantum experiments are conducted using both the IBM Quantum simulator, specifically the \textit{qasm\_simulator}, and an actual IBM quantum computer, specifically the \textit{ibm\_perth}.

The QPT experiments were executed for $11,000$ shots using the \textit{qasm\_simulator} and with $4,000$ repetitions on the real quantum computer \textit{ibm\_perth}. The outcomes were utilized to construct the Choi matrices for the SQSCZ gate. By computing the process fidelity ($0 \leq {\cal F_P} \leq 1$) of noisy quantum channels~\citep{zhang14}, a quantitative comparison was performed to assess the proximity of the measured Choi matrices to the theoretical expectations (ideal process). Figure~\ref{fig:QPTMatrix} illustrates the real and imaginary components of the Choi matrices, along with the corresponding process fidelities derived from the QPT experimental implementations in the two environments. Additionally, for a more detailed analysis of these QPT results, a Hinton diagram depicting the Choi matrices is presented in Figure~\ref{fig:QPTHinton}.

\section{Analysis and Discussions}\label{sec:Analysis}

Quantum computation applications frequently require the utilization of quantum fidelities to assess performance. Quantum fidelity plays a pivotal role in quantifying the similarity between two quantum states. Numerous approaches have been suggested for defining fidelity, typically involving a comparison between the executed gate and the true channel operation. This channel operation can be characterized by the Choi matrix \cite{Choi,Jamiołkowski}.

\medskip

Table~\ref{table:qptSQSCZ} presents detailed device specifications of the \textit{ibm\_perth} and the qubit parameters utilized during the QPT experiments. 
The average coherence times $T1$ and $T2$ are recorded as $118.76681 \mu$s and $131.92139 \mu$s, respectively. 
The average qubit frequency ($\omega/2\pi$) is measured at $5.0676159$ GHz, while 
the average qubit anharmonicity ($\gamma/2\pi$) is determined to be $-0.3419215$ GHz. 
Additionally, the average qubit readout assignment error ($\xi(R)$) is calculated to be $0.02419$. 
The qubit flip probabilities ($\zeta$) and  ($\eta$) exhibit averages of $0.0277333$ and $0.0247666$, respectively. 
The average readout length ($\delta$) is determined to be $721.77777$ ns. 
Figure~\ref{fig:parameters} provides a visualization of the distribution of parameters and readout errors for each qubit across the \textit{ibm\_perth} quantum computer, along with the processor performance during the execution of the QPT experiments.

\medskip

As previously indicated, QPT represents a valuable methodology offering insights into the manner in which a physical process alters states~\cite{qpt,QPT-MaxQ26}. It furnishes detailed insights into the inherent fallibility of processes and the impact of noise or other imperfections on gate operations. Furthermore, QPT serves a crucial function in the validation of quantum gates and the identification of potential shortcomings in their configuration and design\citep{Googles,Sycamore}.

\medskip

The outcomes obtained from the repetition of the quantum experiment for $4,000$ shots for each measurement basis on the \textit{ibm\_perth}, encompassing all the QPT circuits executed for the two-qubit SQSCZ gate, are utilized to reconstruct the Choi matrices. These matrices are depicted in Figures~\ref{fig:QPTMatrix}. 
In Figure~\ref{fig:QPTMatrix}, we present the process matrices illustrating the characteristics of the SQSCZ two-qubit gate in the Pauli basis. Including the theoretically anticipated Choi matrix for the ideal SQSCZ gate, devoid of decoherence effects, with the reconstructed Choi matrices obtained from executing QPT both on a \textit{qasm\_simulator} and an \textit{ibm\_perth} quantum computer. These reconstructions depict both the real ($\mathrm{Re}({\cal C}{\cal E})$) and imaginary ($\mathrm{Im}({\cal C}{\cal E})$) components. Concurrently, Figure~\ref{fig:QPTHinton} showcases the Hinton diagrams, providing further elucidation of the quantum process tomography outcomes. Through the utilization of QPT experiments, we ascertain a process fidelity of $97.27098270\%$ and $88.9938260\%$ for the QPT experiments conducted for the SQSCZ gate using the \textit{qasm\_simulator} and the IBM real quantum machine \textit{ibm\_perth}, respectively.

\begin{table*}
\begin{minipage}{\textwidth}
\caption{\label{table:SQSCZ}The experimental system attributes were observed while executing the experimental implementation of the two-qubit SQSCZ gate for $7,168$ shots on IBM Quantum's quantum computer, named "\textit{ibm\_perth}". The Qiskit runtime time usage amounted to 5 seconds.}
\centering 
\begin{ruledtabular}
\begin{tabular}{ c c c c c c c c c}
$qubit~[i]$ & $T1(\mu s)$ &  $T2(\mu s)$ & $  \omega/2\pi \textrm{(GHz)}$& $\gamma/2\pi \textrm{(GHz)}$&$  \xi (R)$&  $ \zeta (\%)$ &$ \eta (\%)$&$\delta$ (ns)\\
\hline \hline
$qubit~[0]$ &156.94281   &81.627309  &5.157568136   &-0.341524   &0.0280000    &0.0285999   &0.0274   &721.77777   \\
$qubit~[1]$ &89.705850   &47.640857  &5.033544687   &-0.344368   &0.0285999    &0.0288   &0.0284   &721.77777    \\
$qubit~[2]$ &18.130611   &52.050573  &4.862651686   &-0.347272   &0.0282000    &0.0293999   &0.0270   &721.77777    \\
$qubit~[3]$ &198.30965   &170.40818  &5.125102944   &-0.340441   &0.0161000    &0.0172      &0.0150   &721.77777    \\
$qubit~[4]$ &123.95948   &87.855210  &5.159217225   &-0.333366   &0.0276000    &0.0244      &0.0308   &721.77777    \\
$qubit~[5]$ &166.40637   &47.179423  &4.978593106   &-0.346022   &0.0257000    &0.0288  &0.0226   &721.77777    \\
$qubit~[6]$ &201.59061   &246.67977  &5.156639991   &-0.340454   &0.0124999    &0.01319999  &0.0118   &721.77777    \\
\end{tabular}
\end{ruledtabular}
\vspace{0.5em}
\caption{\label{table:qptSQSCZ} The experimental system characteristics were closely monitored duringthe execution of the Quantum Process Tomography (QPT) experiments for the 2-qubit SQSCZ gate. The QPT experiment was iterated 4,000 times, employing IBM Quantum's quantum computer, \textit{ibm\_perth}. The total runtime amounted to 2 minutes and 40 seconds.}
\centering 
\begin{ruledtabular}
\begin{tabular}{ c c c c c c c c c}
$qubit~[i]$ & $T1(\mu s)$ &  $T2(\mu s)$ & $  \omega/2\pi \textrm{(GHz)}$& $\gamma/2\pi \textrm{(GHz)}$&$  \xi (R)$&  $ \zeta (\%)$ &$ \eta (\%)$&$ \delta$ (ns)\\
\hline \hline
$qubit~[0]$ &130.04457328   &73.58860043  &5.1575647635   &-0.341524517   &0.028399999    &0.029399999   &0.0274   &721.77777    \\
$qubit~[1]$ &139.75002257   &54.06849027  &5.0335394509   &-0.344368702   &0.028499999    &0.027         &0.03     &721.77777    \\
$qubit~[2]$ &20.132049776   &75.47632710  &4.8626527977   &-0.347272472   &0.027399999    &0.031399999   &0.0234   &721.77777    \\
$qubit~[3]$ &181.07440147   &194.4456264  &5.1250989116   &-0.340441899   &0.016999999    &0.0194        &0.014599999   &721.77777    \\
$qubit~[4]$ &133.35414902   &118.6893525  &5.1592189326   &-0.333366537   &0.028399999    &0.0292        &0.027599999   &721.77777    \\
$qubit~[5]$ &80.657491205   &176.7236497  &4.9785968466   &-0.346022031   &0.027800000    &0.0300        &0.0256   &721.77777    \\
$qubit~[6]$ &146.35500245   &230.4577368  &5.1566401889   &-0.340454390   &0.011900000    &0.0153999999  &0.0084   &721.77777    \\
\end{tabular}
\end{ruledtabular}
\end{minipage}
\end{table*}

\section{Conclusion}\label{SEC:Concl}

In summary, quantum process tomography  is an essential procedure aimed at validating quantum gates and identifying shortcomings in configurations and gate designs. Its recent utilization has played a crucial role in experimental quantum computation and communication. 
In this study, we assess the SQSCZ gate's performance on a noisy intermediate-scale quantum computer through QPT experiments conducted across diverse environments, employing both IBM Quantum's simulators (\textit{qasm\_simulator}) and a real quantum computer (\textit{ibm\_perth}). 
The QPT experiments were conducted with 11,000 repetitions for each measurement basis using the \textit{qasm\_simulator}, and with 4,000 shots on the \textit{ibm\_perth}. The resulting measurement outcomes from all circuits were utilized to reconstruct the Choi matrices, which were then compared with ideal expectations. Our findings reveal comparable fidelities between theoretical predictions and experimental implementations, shedding light on the accuracy of implementations, measurement errors, and noise properties of the SQSCZ gate. Specifically, we demonstrated a process fidelity (${\cal F_P}$) of $97.27098\%$ and $88.99383\%$, respectively. 
These findings bear promising implications for advancing both theoretical understanding and practical applications within the domain of quantum computation.

\section{List of Abbreviations }

\begin{table}[H]
    \centering 
    \begin{tabular}{ll}
    NISQ         &Noisy intermediate-scale quantum    \\
    QPT          &Quantum process tomography    \\
    QST          &Quantum state tomography     \\
    ${\cal P}_{\rm{in(out)}}$   &Input (output) density matrix \\
    ${\cal E}$   &A quantum dynamical map (process)\\
    $\chi_{mn}$ &A positive Hermitian matrix \\
    $\widehat{W}_m$ &Operators constitute a complete basis\\
    CPTP         &Completely positive and trace-preserving\\
    PTM          &Pauli-Transfer-Matrix\\
    $(.)^T$           &The matrix transposition\\
    ${\cal C}_{\cal E}$ &The Choi matrix \\
    ${\cal H}_\mathrm{in(out)}$ &The input (output) Hilbert spaces\\
    $\Xi$        & Projector \\
    $\textrm{Tr}^\pounds$ &The partial trace\\
    ID           &Identitiy gate\\
    SX           &The square root of X gate\\
    $R_Z$        &The rotation about $z$-axis\\
    Ph           & Phase gate\\
    CZ           &The controlled-Z gate \\
    CX           &The controlled-X gate \\
    $\sqrt{CZ}$  &The square root of controlled-Z\\
    $\sqrt{SWAP}$&The square root of SWAP gate\\
    QFT          &Quantum Fourier transform \\
    Falcon r5.11H&A quantum processor by IBM Quantum\\
    QV           &Quantum voulume\\
    ${\cal F_P}$ &The process fidelity \\
    qasm            &Quantum assembly language \\
    $qubit~[i]$     &The $i^{th}$  qubit \\
    $T1$            &Relaxation time    \\ 
    $T2$            &Dephasing time    \\ 
        \end{tabular}
     \end{table}  
\begin{table}[H]
\centering 
\begin{tabular}{lll}
    $\omega/2\pi$   &&The qubit frequency    \\ 
    $\gamma/2\pi $  &&The qubit anharmonicity    \\ 
    $\xi(R)$        &&The qubit readout assignment error    \\ 
    $ \zeta $       &&The Probability of meas. 0 prep. 1   \\ 
    $ \eta $        &&The Probability of meas. 1 prep. 0    \\ 
    Re[.]           &&Real part \\
    Im[.]           &&Imaginary Part \\
    GHz             &&Giga hertz   \\%$1 \mathrm{ GHz} =1.0\times 10^{9}$ Hertz  \\ 
    MHz             &&Mega hertz   \\%$1 \mathrm{ MHz} =1.0\times 10^{6}$ Hertz \\ 
    $\mu$s          &&Micro seconds \\%, 1 $\mu \mathrm{s}=1.0\times 10^{-6}$ second \\
    ns &&Nano seconds \\%, 1 $ \mathrm{ns}=1.0\times 10^{-9}$ second \\
    \end{tabular}
\end{table}

\begin{acknowledgments}

``We acknowledge the use of IBM Quantum for this research. The views and conclusions expressed by the author are their own and do not necessarily reflect IBM Quantum’s official policy or position. 
This research received no specific grant from any funding agency in the public, commercial, or not-for-profit sectors.” 

\end{acknowledgments}

% \section*{Authorship contribution statement}

% M. AbuGhanem: Conceptualization,  Methodology, Resources, Data curation, Formal analysis, Software, Visualization, Investigation, Validation, Writing, review  and editing. 

\section*{Declaration of competing interest}

``The author declare that they have no known competing financial interests or personal relationships that could have appeared to influence the work reported in this paper."

\section*{Data availability}

The data sets produced and/or analyzed during the current study are incorporated within this article.

\end{document}